\documentclass[journal,comsoc]{IEEEtran}
\usepackage[T1]{fontenc}

\ifCLASSINFOpdf
   \usepackage[pdftex]{graphicx}
  \DeclareGraphicsExtensions{.pdf,.jpeg,.png}
\else
\fi
\usepackage{amsmath}
\usepackage[cmintegrals]{newtxmath}
\usepackage[caption=false]{subfig}
\usepackage{mathtools, cuted}

\usepackage{graphicx}
\usepackage{latexsym}
\usepackage{amsmath,epsfig,mathrsfs}
\usepackage{epstopdf}
\usepackage{multirow}
\usepackage{subfig}
\usepackage{booktabs,siunitx}
\usepackage{lscape}
\usepackage{url}
\usepackage{graphicx}
\usepackage{epsfig}
\usepackage{graphicx}
\usepackage{latexsym}
\usepackage{amsmath}
\usepackage{amssymb}
\usepackage{mhsetup}
\usepackage{mathtools}
\usepackage{mathrsfs}
\usepackage{float}
\usepackage{xfrac}
\usepackage{epsfig}
\usepackage[english]{babel}
\usepackage{caption}

\begin{document}

\title{Performance Analysis of an Outdoor Visible Light Communications System for Internet-of-Vehicles}

\author{Ramadan~Abdul-Rashid, Anas M. Salhab and Salam A. Zummo\\
	
	Electrical Engineering Department, King Fahd University of Petroleum and Minerals, Dhahran, Saudi Arabia \\
	Email: {g201409740@kfupm.edu.sa,salhab@kfupm.edu.sa, zummo@kfupm.edu.sa}}
 
\maketitle

\begin{abstract}
This paper looks at the performance of an outdoor visible light (OVLC) communication system used for Internet-of-Vehicles (IoV). In the proposed system an amplify-and-forward (AF) opportunistic scheme is used to extend the range of information broadcast from a traffic light to vehicles. The Gamma-Gamma channel gain and the Lambertian direct current (DC) channel gain are used to model the fading coefficient of each transmission and the  short and thermal noise models used to represent the system noise. The statistics of the system are examined by deriving closed-form expressions for the cumulative distribution function (CDF) and probability density function (PDF) for the equivalent end-to-end signal-to-noise ratio (SNR). Novel closed form equations are also developed for the outage probability and ergodic capacity. Numerical simulations performed showed that the system performance in terms of both outage probability and ergodic capacity improves with decreasing turbulence intensity. Results also illustrate that, as the distance between the vehicles increases, the system performance is more deteriorated.
\end{abstract}


\IEEEpeerreviewmaketitle

\section{Introduction}
The  advent of high speed Internet and the rapidly evolving fields of artificial intelligence (AI) and intelligent transportation systems (ITS) has led to the development of a special area of Internet-of- Things (IoT) known as the  Internet-of-Vehicles (IoV). With the development of highly efficient systems of computing and wireless communications, IoV offers a myriad of potential applications in both civil and military sectors \cite{cheng2015routing}. Some of such applications span from communication of military vehicles on the field to traffic light broadcast communications used for road safety and traffic control. These applications among others have led to an increased interest in the research, design and deployment of IoV systems. Most reported applications for IoV communicate in the radio frequency (RF) domain \cite{hamid2019internet} where there is limited bandwidth allocated for communications.  This weakness allows for a high interference among several vehicles within the same vicinity and attempting to communicate concurrently. This flaw calls for a new method for vehicles to communicate in the wireless domain. Visible light communication (VLC), is a relatively new method for wireless communications which is well suited for such applications as IoV since it is well known for low cost high speed wireless applications \cite{ndjiongue2018overview}. 

Recent advances in materials and solid-state engineering have
enabled the development of highly efficient LEDs that are now
being widely used in outdoor lighting such as, traffic lights, bill-boards, street lights, vehicular lightning, and advertising displays. Also, several vehicular manufacturers have switched from traditional lightning techniques to LED-based lightning in the production of vehicles \cite{lourencco2012visible}. Due to the rampant adoption of LEDs for both
in indoor and outdoor lightings, the VLC becomes a promising candidate for future ITSs. Up to now, the VLC capabilities of delivering high data rate has been investigated for both the indoor and outdoor environments. For the indoor environment, most works focus on channel modeling, resource allocation and performance analysis for VLC systems \cite{sun2017superimposed}. In the outdoor environment, most works have focused on channel and noise modeling, modulation methods, multiplexing techniques, pre-equalization and post-equalization schemes, diversity reception schemes, and general performance analysis. In \cite{lee2009performance}, the authors propose a day-light noise model and a novel receiver model that employs selection combining that significantly minimizes the effects of background noise and showed that, their developed system achieved stable communication whiles improving signal-to-noise ratio (SNR). Kim \textit{et al} \cite{kim2012outdoor}, presented an experimental approach to study the effect of day-light noise on outdoor VLCs. They purported the presence of saturation in photo detectors which deteriorates system performance. In \cite{luo2015performance}, the performance of an inter-vehicular outdoor VLC system in terms of bit error rate (BER) and communication range was presented. In their systems model, the authors considered both line-of-sight (LOS) and non-LOS signals. They also showed that, when a photodetector is installed above 200 cm, coverage of communication can reach up to around 70 m and achieve rates of around $50$ Mb/s. They also showed that, a wet road can improve the BER performance. 

In an outdoor scenario, although misalignment fading has less effect on outdoor visible light communication (OVLC) due to the large angle of emergence of LED, atmospheric channel fading is a crucial factor that degrades performance severely \cite{ndjiongue2018overview}. The atmospheric channel fading are characterized as strong and weak based on the distance between the transmitter and receiver. Different atmospheric turbulence models have been proposed for outdoor optical wireless communications. The channel models normally utilized are the negative exponential model which represents strong turbulence, the log-normal model which is used to model weak turbulence and the Gamma-Gamma model which can represent weak, moderate and strong atmospheric turbulences.

In \cite{sun2017superimposed}, a new relaying scheme was proposed for outdoor inter-vehicular OVLC system. The proposed method called superimposed relaying was used to send information broadcast from a traffic light to a car by another car acting as a relay. The log-normal atmospheric turbulence model was used to model the channel gain. They derived a closed form expression for the BER and used it to optimize the power allocated to each signal used for transmission. Lin \textit{et al.} \cite{lin2017outage} also looked at the outage performance analysis of an outdoor vehicle-to-vehicle VLC system (V2LC). A lognormal atmospheric turbulence model was used to derive a closed-form expression for the outage probability. They also asserted that, the distortions to the system as mainly due to short noise from background noise and thermal noise from the receiver electronics. The derived outage model was shown to accurately evaluate the system performance. Although, these works have addressed several design issues pertaining to OVLC, the performances are limited to outage and BER. Also, the model used for the atmospheric channel gain is the lognormal which only approximate weak turbulence. 

The Gamma-Gamma distribution is the most preferred model for atmospheric channel fading since it covers both weak and strong turbulence \cite{salhab2016power}. As an effective method to bypass the obstacle and mitigate fading, relaying techniques have attracted significant attention in recent literature. With the extensive investigation of cooperative relaying systems, there are several cooperative protocols which have been researched. Decode-and-forward (DF) and amplify-and-forward (AF) are two common protocols \cite{sun2017superimposed} employed in cooperative scheduling.  

In this work we evaluate the performance of an OVLC with an AF relay in which the relay node forwards information from the source to a destination using visible light communications whiles considering a Gamma-Gamma atmospheric model and also the Lambertian DC gain. This system is motivated by IoV where a vehicle communicates with a traffic light and also acts as a relay node to extend coverage to another vehicle not in LOS to the traffic light. To the best of our knowledge, the outage probability and the ergodic capacity of such a system has not yet been investigated.  According to  \cite{lin2017outage}, by considering the impacts of path loss and atmosphere turbulence, a statistical channel model can be established for this system and the outage probability can be developed.

The key contributions of this paper are outlined below.
\begin{enumerate}
	
	\item Establish a statistical model for the end-to-end equivalent SNR for a dual-hop AF OVLC network with Gamma-Gamma atmospheric turbulence which includes the cumulative distribution function (CDF) and probability density function (PDF) .
	\item Derive closed for expressions for the outage probability and ergodic capacity of the system.
	\item Conduct numerical simulations for outage probability and ergodic capacity for weak, moderate, and strong atmospheric turbulences. 
\end{enumerate}

The rest of the paper is organized as follows. Sections II provides the system and channel models. Section III presents the statistics of the equivalent SNR of the system. The performance analysis of the system in terms of outage probability and ergodic capacity is presented in Section IV. Section V discusses some numerical results. The paper is then concluded in Section VI.


\section{System Model}
\begin{figure*}[ht] 
	\centering    
	\includegraphics[width=0.7\textwidth]{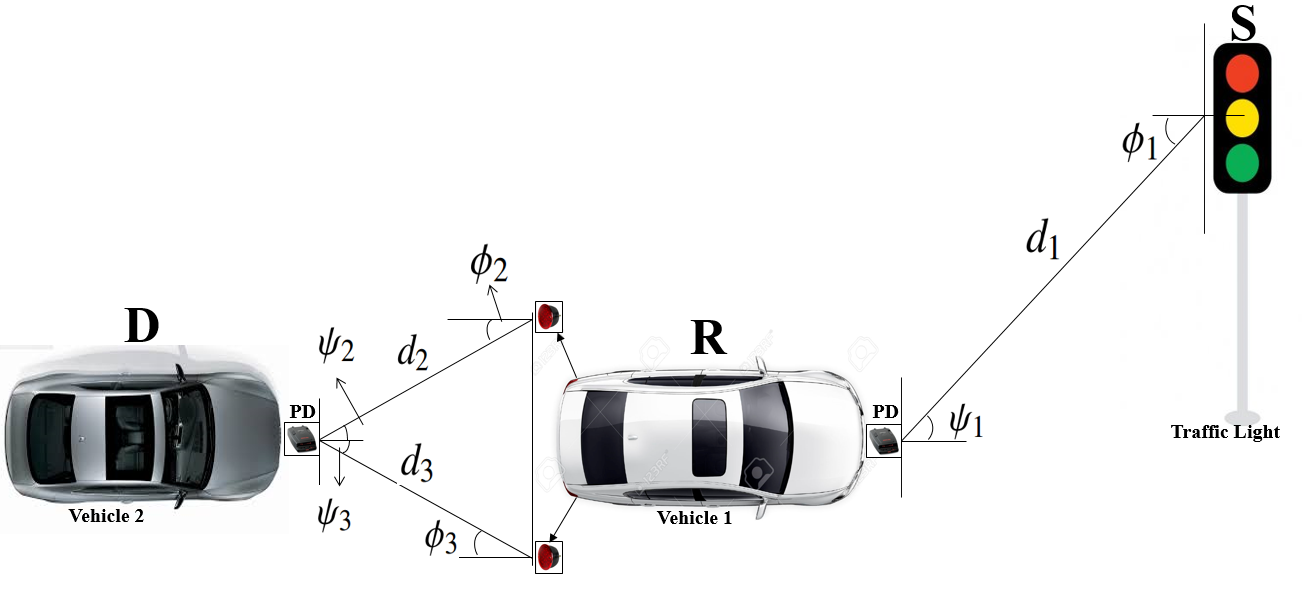}
	\caption[An OVLC cooperative system model for Internet of Vehicles]{An OVLC cooperative system model for Internet of Vehicles}
	\label{fig:model}
\end{figure*}

We consider an opportunistic OVLC communication system with three nodes as shown in \ref{fig:model}. This transmission process is composed of two stages. In the first stage the traffic light which represents the source node (S) communicates with Vehicle 1 which acts as the relay (R). The   received signal at R can be given by:

\begin{equation}\label{eq1}
y_r = g_s h_{sr} x_s + z_{sr},
\end{equation}

where $g_s$ and  $x_s \geq 0$ denote the optoelectronic conversion factor and the non-negative visible-light signal from the LEDs on traffic light. $z_{sr}$ is an additive white Gaussian noise with zero mean and variance $\sigma_{z_1}^{2}$.

The channel fading coefficient is composed of both the DC path loss based on the Lambertian  emission $h^{Lm}_{sr}$ and the atmosphere turbulence, which describes the effects of atmospheric weather and environment conditions, $h^{Tu}_{sr}$ and related to the overall channel fading coefficient by:

\begin{equation}\label{eq2}
h_{sr} = h^{Lm}_{sr}h^{Tu}_{sr}.
\end{equation}

In the second stage, R transmits an amplified forms of $x_s$ from its tail-lights to the Vehicle 2 which serves as the destination node (D). Since two signals are arriving at the PD receiver on D, the received signal(s) can be represented in a vector form as:

\begin{equation}\label{eq3}
\textbf{y}_d = g_r \textbf{h}_{rd} x_r + \textbf{z}_{rd},
\end{equation}

where $\textbf{h}_{rd} = [h_{rd,1},h_{rd,2}]^T$, $h_{rd} = h^{Lm}_{rd} h^{Tu}_{rd}$, $\textbf{y}_d = [y_{rd,1},y_{rd,2}]^T$, and  $\textbf{z}_{rd} = [z_{rd,1},z_{rd,2}]^T$

The transmitted signal from the relay, $x_r$ is related to $y_{sr}$ by:

\begin{equation}
x_r = G y_{sr}
\end{equation} 

where $G$ is the relay gain chosen as \cite{salhab2016power}, $G = (g_r |h_{sr}|^2 + \sigma_{z_1}^2 )^{-\frac{1}{2}}$.

\section{Channel Fading Model}
The Lambertian DC chhannel gain can be expressed as \cite{ndjiongue2018overview}:

\begin{equation}
h^{Lm}_i =
\begin{cases}
\frac{(m+1)A}{2 \pi d_i} cos^m(\phi_i) cos(\psi_i) T_f(\psi_i)T_c(\psi_i), & 0 \leq \psi\leq \Psi, \\
0, & \psi > \Psi,
\end{cases}
\end{equation}

\noindent
where $m$ is the order of the Lambertian emission, $A$ is the
receiver area of the PD, assuming same $A$ for  both PDs, $d_i$, $\phi_i$ and $\psi_i$ are the distance, the
angle of irradiance and the angle of incidence from the $i$-th
LED to the PD, respectively. $T_f(\phi_i)$ and $T_c(\phi_i)$ are the optical
filter gain and the concentrator gain of the PD, respectively, and $\Psi$ is the field-of-view (FoV) of the PD.

Due to the wide FoV of the VLC signal, it is fair to assume perfect alignment between the VLC transmitter and receiver apertures, and considering Gamma-Gamma turbulence-induced fading, fluctuations in the received optical irradiance, the PDF of the turbulence induced fading coefficient for the $i$-th link, $h^{Tu}_i$ can be given as:

\begin{equation}
f_{h^{Tu}_i}(h^{Tu}_i)=\frac{{2(\alpha\beta)}^\frac{\alpha+\beta}{2}}{\Gamma(\alpha)\Gamma(\beta)} (h^{Tu}_i)^{\frac{\alpha+\beta}{2}-1}K_{\alpha-\beta}\left(2\sqrt{\alpha\beta h^{Tu}_i}\right),\quad h^{Tu}_i\ge 0.
\end{equation}

The instantaneous received electrical SNR, of the $i$-th VLC link is related to $h^{Tu}_i$ as
$\gamma_i = \bar{\gamma}_i |h^{Tu}_i|^2$, where $\bar{\gamma}_i$ is the average electrical SNR which is defined as $\bar{\gamma}_i = \frac{(g \times h^{Lm}_i P_i)^2}{\sigma^2_{z_i}}$, where $g$ denotes the optoelectronic conversion factor. For  non-negative visible-light signal from a VLC transmitter to a PD receiver of the $i$-th link, $x_i \geq 0$, $P_i = \mathbb{E}[x_i]$ is the transmitted average optical intensity for the $i$-th VLC link and $\sigma^2_{z_i}$ is the log-amplitude variance, where $\mathbb{E}[.]$ denote an expectation operation. Using power transformation of random variables, it is easy to show that the PDF of $\gamma_i$ is given by:

\begin{align}\label{pdfSNR}
f_{\gamma_{i}}(\gamma)=\frac{{\left(\alpha\beta/\sqrt{\bar{\gamma}_{i}}\right)} ^\frac{\alpha+\beta}{2}}{\Gamma(\alpha)\Gamma(\beta)}{\gamma}^{\frac{\alpha+\beta}{4}-1}\times K_{\alpha-\beta} \left(2\sqrt{\frac{\alpha\beta}{\sqrt{\bar{\gamma}_{j}}}{\gamma}^{\frac{1}{2}}}\right), {\gamma_{i}} \ge 0.
\end{align}\\

The parameters $\alpha$ and $\beta$ are used to vary the conditions of the atmospheric turbulence, and the expressions  $K_v(.)$ is the $ν-th$ order modified Bessel function of the second kind defined in \cite{korenev2003bessel}. Normally, the parameters $\alpha$ and $\beta$ represent the effective quantity of small-scale and large-scale eddies of the turbulent environment, respectively. Based on the values of these parameters, the turbulent component of the channel gain can be modeled within strong, medium and weak regimes. The relations of $\alpha$ and $\beta$ for different propagation situations are given by \cite{salhab2016power}. Assuming a spherical wave propagation, $\alpha$ and $\beta$ can be defined as \cite{salhab2016power}: 

\begin{align} \alpha &=\left[\exp \left\lbrace \frac{0.49\kappa ^{2}}{\left[1+0.18\rho ^{2}+0.56\kappa ^{12/5}\right]^{7/6}}\right\rbrace -1\right]^{-1},\\ \beta &=\left[\exp \left\lbrace \frac{0.51\kappa ^{2}[1+0.69\kappa ^{12/5}]^{-5/6}}{\left[1+0.9\rho ^{2}+0.62\xi ^{2}\kappa ^{12/5}\right]^{5/6}}\right\rbrace -1\right]^{-1}, \end{align}

\noindent
where $\rho$ is the optical wave number given in terms of the refractive index parameter, $C^2_n$, the VLC signal wavelength, $\lambda_{VLC}$ and the diameter of the receiver aperture $D$ as $\rho = \sqrt{\frac{2\pi k D^2}{4 \lambda_{VLC}} d_i}$. Also, $\kappa = 0.5d^{11/6}_i C^2_n (\frac{2\pi}{\lambda_{VLC}})$ is the Rytov variance.

\section{Noise Model}
The noise process in an outdoor VLC, is mainly due two noise components. These are the ambient-induced shot noise and the thermal noise. For outdoor VLC, the the communication signals are highly affected by the visible background day-time light. This background daylight leads to the an induction of large amplitude short noise in the PD of the receiver \cite{lin2017outage}. At the receiver, an optical filter can be used to minimize the background light. Both the shot noise and the thermal noise can be modeled using the Gaussian distribution. Say $\sigma^2_{z_i}$ is the variance of the noise process of $z_i$ for the $i$-th VLC link, which can be expressed as the sum of the shot noise, $\sigma^2_{i,shot}$ and the thermal noise, $\sigma^2_{i,thermal}$ as follows  \cite{lin2017outage},

\begin{equation*} \sigma_{z_i}^{2}=\sigma_{i,\text{shot}}^{2}+\sigma_{\text{i,thermal}}^{2}. \tag{6} \end{equation*}

The shot noise can also be expressed as \cite{lee2009performance}:

\begin{equation*}\sigma_{i,shot}^{2}= 2gQ W_n[P_ih_i + W_R \xi T_o A T_c sin^2(\xi)], \end{equation*}

\noindent
where, $Q$ is the electronic charge, $W_n$ is the equivalent noise
bandwidth, $W_R$ is the noise bandwidth factor for a rectangular
transmitter pulse shape, $T_0$ is the peak filter transmission coefficient, $n_i$ denotes the internal refractive index of the optical concentrator, $\xi$ is the irradiance that falls within the spectral range of the receiver, i.e., \cite{lee2009performance}

\begin{equation*} \xi = \int\nolimits_{\l_{1}}^{l_{2}}S_{\text{peak}}\frac{\chi (\lambda, T_{\mathrm{B}})}{\max\limits_{l}\chi(\lambda, T_{\mathrm{B}})}\mathrm{d}l, \tag{10} \end{equation*}

\noindent
where $l_{1}$ and $l_{2}$ (in $\mu m$) are the lower and upper spectral limits of the optical bandpass filter at the receiver respectively, $S_{\text{peak}}$ (in $W/m^2$) denotes peak spectral irradiance, $\chi (\lambda, T_B)$ is the spectral irradiance of the Blackbody radiation model, which can be expressed as:

\begin{equation*} \chi (\lambda, T_{\mathrm{B}})=\frac{2\pi \nu c^{2}}{\lambda^{5}(e^{\frac{\nu c}{\lambda kT_{\mathrm{B}}}}-1)}, \tag{11} \end{equation*}

where $\nu$ denotes the Planck's constant, $c$ is the speed of light,
$\lambda$ is the wavelength, $k$ is the Boltzman's constant, and $T_B$ (in
Kelvin) is the average surface temperature of the sun.

Likewise, the thermal noise has two components,  include two parts: the
feedback-resistor noise and the FET channel noise. Therefore,  $\sigma_{\text{i,thermal}}$
is given by:

\begin{equation*} \sigma_{\text{Thermal}}^{2}=\underbrace{\frac{8\pi kT_{a}}{G}\eta AW_{R}^{2}}_{\text{feedback}-\text{resistor noise}}+\underbrace{\frac{16\pi^{2}kT_{a}\Omega}{g_{m}}\eta^{2}A^{2}W_{RC}W^{3}_R}_{\text{FET channel noise}}, \tag{12} \end{equation*}

\noindent
where $T_a$ is the absolute temperature, $G$ is the open-loop
voltage gain, $\eta$ is the fixed capacitance per unit area, $\Omega$ is the
FET channel noise factor, $g_m$ is the FET transconductance, and $W_{RC}$
is the noise bandwidth factor for a full raised-cosine equalized
pulse shape.

\section{Statistics of the Equivalent SNR at Destination}
With the signal model given by (\ref{eq1}), (\ref{eq2}), and (\ref{eq3}), the effective end-to-end SNR at the destination node, $\gamma_{\sf D}$ can  be derived in terms of the SNR between the source (Traffic light) and the relay node (Vehicle 1), $\gamma_{{{\sf sr}}}$ and the effective SNR between the relay and the destination (Vehicle 2), $\gamma_{{{\sf rd}}}$ as \cite{salhab2016power}:

\begin{align}\label{eqSNR}
\gamma _{\sf D}=\frac{\gamma _{{\sf sr}}\gamma _{\sf rd}}{\gamma _{{\sf sr}}+\gamma _{\sf rd}+1}. 
\end{align}

\noindent
Since two links exist between the relay node and the destination node, we employ selection combining (SC) at the destination to determine the SNR in this link, which is given by \begin{align} \gamma _{\sf rd} = \text{max} \{ \gamma _{{\sf rd,1}},\gamma _{\sf rd,2} \}. \end{align} 
Using the standard approximation, $ \gamma _{\sf D} \cong \text{min} \{ \gamma _{{\sf sr}},\gamma _{\sf rd} \}$, the SNR in (\ref{eqSNR}) can be upper bounded \cite{salhab2016power}: 

\begin{align}\label{SNRA} 
\gamma _{\sf D}\leq \frac{\gamma _{{\sf sr}}\gamma _{\sf rd}}{\gamma _{{\sf sr}} + \gamma _{\sf rd}}. \end{align}

\subsection{Cumulative Distribution Function (CDF)}
The effective SNR given in (\ref{SNRA}) is shown to have a CDF as follows
\begin{align} \label{eqssnr}
F_{\gamma _{\sf d}}(\gamma) = F_{\gamma _{\sf sr}}(\gamma)+F_{\gamma _{\sf rd}}(\gamma)-F_{\gamma _{\sf sr}}(\gamma)F_{\gamma _{\sf rd}}(\gamma). \end{align}

And by employing SC at D, the CDF of $\gamma_{{{\it rd}}}$ is written in terms of the CDFs of the SNRs of the links between the relay and the destination, $\gamma_{{\sf rd,1}}$ and $\gamma_{\sf rd,2}$ as

\begin{align} \label{eqSNRRD}
F_{\gamma _{\sf rd}}(\gamma) = F_{\gamma _{\sf rd,1}}(\gamma)F_{\gamma _{\sf rd,2}}(\gamma), \end{align}

Combining (\ref{eqssnr}) and (\ref{eqSNRRD}), we obtain

\begin{align}\label{cdfD}
F_{\gamma _{\sf d}}(\gamma) = F_{\gamma _{\sf sr}}(\gamma) + F_{\gamma _{\sf sr,1}}(\gamma) F_{\gamma _{\sf rd,2}}(\gamma) - F_{\gamma _{\sf sr}}(\gamma) F_{\gamma _{\sf rd,1}}(\gamma)F_{\gamma _{\sf rd,2}}(\gamma), \end{align}

Given the PDF, $f_{\gamma_{{{\sf i}}}} (\gamma_{{{\sf }}})$ of the SNR, $\gamma_{{{\sf i}}}$, it is well known that $F_{\gamma_{{{\sf i}}}} (\gamma_{{{\sf }}})$ can be obtained by integrating the PDF within the limits of the SNR, i.e.,

\begin{align} F_{\gamma _{\sf i}}(\gamma)=\int _{0}^{\gamma }f_{\gamma _{\sf i}}(\zeta)d\zeta. \end{align}

Thus we derive the CDF of $\gamma_{{{\sf i}}}$ by integrating (\ref{pdfSNR}), which is given by:

\begin{equation}\label{cdfSNR}
F_{\gamma_{i}}(\gamma) = \frac{4{\left(\alpha\beta/\sqrt{\bar{\gamma}_{i}}\right)} ^\frac{\alpha+\beta}{2}}{(\alpha+\beta) \Gamma(\alpha)\Gamma(\beta)}{\gamma}^{\frac{\alpha+\beta}{4}-1}\\\times K_{\alpha-\beta} \left(2\sqrt{\frac{\sqrt{\alpha\beta \gamma}}{\sqrt{\bar{\gamma}_i}}} \right), {\gamma} \ge 0.
\end{equation}

Due to the large spacing between the two LEDs of Vehicle 2, we assume independence between their respective SNRs, $\gamma_{{\sf rd,1}}$ and $\gamma_{{\sf rd,2}}$. Hence the effective SNR between the relay and destination nodes $\gamma_{{\sf rd}}$, can be given by:

\begin{align} F_{\gamma _{\sf rd}}(\gamma)=\prod _{k=1}^{2}\int _{0}^{\gamma }f_{\gamma _{\sf rd,k}}(\zeta)d\zeta, \end{align}

The CDF of SNR of the effective end-to-end for the OVLC link in our model is thus obtained by calculating the CDFs for each link and substituting the results in (\ref{cdfD}). After some manipulations, a closed-form expression is obtained for the CDF, $F_{\gamma_{D}}(\gamma)$ as:

\begin{strip}
\begin{align} \label{CDF}
F_{\gamma_{D}}(\gamma)   =&   4\,{\frac {{\gamma}^{\alpha/4+\beta/4 - 1}}{ \left( \alpha +\beta \right) \Gamma
		\left( \alpha \right) \Gamma \left( \beta \right) } \left( {\frac {\alpha 
			\beta }{\sqrt {\bar{\gamma}_{{{\sf sr}}}}}} \right) ^{\alpha/2+\beta/2}{
		{\sl K}_{\alpha-\beta}\left(2\,\sqrt {{\frac {\sqrt {\alpha \beta \,
						\gamma}}{\sqrt {\bar{\gamma}_{{{\sf sr}}}}}}}\right)}} \nonumber\\ &+ 16\,{\frac { \left( {
			\gamma}^{(\alpha +\beta)/2} \right) ^{2}}{ \left( \alpha +\beta \right) ^{2}
		\left( \Gamma \left( \alpha  \right)  \right) ^{2} \left( \Gamma \left( \beta
		\right)  \right) ^{2}} \left(  \left( {\frac {\alpha \beta }{\sqrt {
				\bar{\gamma}_{{{\sf rd}}}}}} \right) ^{\alpha/2+\beta/2} \right) ^{2} 
	\left( {{\sl K}_{\alpha-\beta}\left(2\,\sqrt {{\frac {\sqrt {\alpha 
						\beta \,\gamma}}{\sqrt {\bar{\gamma}_{{{\sf rd}}}}}}}\right)} \right) ^{2}} \\ \nonumber&- 
64\,{\frac { \left( {\gamma}^{3\alpha/4+3\beta/4} \right)}{ \left( 
		\alpha +\beta \right) ^{3} \left( \Gamma \left( \alpha \right)  \right) ^{3} \left( 
		\Gamma \left( \beta \right)  \right) ^{3}} \left( {\frac {\alpha \beta }{
			\sqrt {\bar{\gamma}_{{{\sf sr}}}}}} \right) ^{\alpha/2+\beta/2}{{\sl K}_{
			\alpha-\beta}\left(2\,\sqrt {{\frac {\sqrt {\alpha \beta \,\gamma}}{
					\sqrt {\bar{\gamma}_{{{\sf sr}}}}}}}\right)} \left(  \left( {\frac {\alpha 
			\beta }{\sqrt {\bar{\gamma}_{{{\sf rd}}}}}} \right) ^{\alpha/2+\beta/2}
	\right) ^{2} \left( {{\sl K}_{\alpha-\beta}\left(2\,\sqrt {{\frac {
					\sqrt {\alpha \beta \,\gamma}}{\sqrt {\bar{\gamma}_{{{\sf rd}}}}}}}\right)}
	\right) ^{2}}
\end{align}
\end{strip}

\begin{strip}
	\begin{align}\label{PDF}
 f_{\gamma_{D}}(\gamma) =& \,{\frac {{\gamma}^{(\alpha+\beta)/4 -1}}{\Gamma
			\left( \alpha \right) \Gamma \left( \beta \right) } \left( {\frac {\alpha 
				\beta }{\sqrt {\bar{\gamma}_{{{\sf sr}}}}}} \right) ^{(\alpha+\beta)/2}{
			{\sl K}_{\alpha-\beta}\left(2\,\sqrt {{\frac {\sqrt {\alpha \beta \,
							\gamma}}{\sqrt {\bar{\gamma}_{{{\sf sr}}}}}}}\right)}} \nonumber + 8\,{\frac { \left( {
				\gamma}^{(\alpha +\beta)/2 - 1} \right) ^{2}}{ \left(\alpha +\beta \right) 
			\left( \Gamma \left( \alpha  \right)  \right) ^{2} \left( \Gamma \left( \beta
			\right)  \right) ^{2}} \left(  \left( {\frac {\alpha \beta }{\sqrt {
					\bar{\gamma}_{{{\sf rd}}}}}} \right) ^{(\alpha +\beta)/2} \right) ^{2} 
		\left( {{\sl K}_{\alpha-\beta}\left(2\,\sqrt {{\frac {\sqrt {\alpha 
							\beta \,\gamma}}{\sqrt {\bar{\gamma}_{{{\sf rd}}}}}}}\right)} \right) ^{2}} \nonumber\\ &- 
	48\,{\frac { \left( {\gamma}^{3(\alpha+\beta)/4 - 1} \right)}{ \left( 
			\alpha +\beta \right) ^{2} \left( \Gamma \left( \alpha \right)  \right) ^{3} \left( 
			\Gamma \left( \beta \right)  \right) ^{3}} \left( {\frac {\alpha \beta }{
				\sqrt {\bar{\gamma}_{{{\sf sr}}}}}} \right) ^{(\alpha+\beta)/2}{{\sl K}_{
				\alpha-\beta}\left(2\,\sqrt {{\frac {\sqrt {\alpha \beta \,\gamma}}{
						\sqrt {\bar{\gamma}_{{{\it sr}}}}}}}\right)} \left(  \left( {\frac {\alpha 
				\beta }{\sqrt {\bar{\gamma}_{{{\sf rd}}}}}} \right) ^{(\alpha+\beta)/2}
		\right) ^{2} \left( {{\sl K}_{\alpha-\beta}\left(2\,\sqrt {{\frac {
						\sqrt {\alpha \beta \,\gamma}}{\sqrt {\bar{\gamma}_{{{\sf rd}}}}}}}\right)} \right) ^{2}}
	\end{align}
\end{strip}

\subsection{Probability Density Function (PDF)}
Utilizing the closed-form expression of the CDF in (\ref{CDF}), the PDF of the equivalent end-to-end SNR,$f_{\gamma_{D}}(\gamma)$  is obtained by differentiating (\ref{CDF}) with respect to $\gamma_{{{\it }}}$ as shown in (\ref{diff}). This yields a the novel closed-form PDF given by (\ref{PDF}).

\begin{equation}\label{diff}
f_{\gamma_{\sf D}} (\gamma)  =  \frac{d}{d \gamma} F_{\gamma_{D}}( \gamma)
\end{equation}

\section{Performance Evaluation}
In this section, the performance of the OVLC model is evaluated using the metrics of outage probability and ergodic capacity. Closed-form expressions for each of these metrics are presented.

\subsection{Outage Probability}
In wireless networks, an outage event in communication is said to have occured if the SNR at a certain time instant of a link falls below some pre-determined threshold ($\gamma_{{{\it \text{out}}}}$). Therefore we can define the probability of outage as the probability that the effective end-to-end SNR at the destination node falls below $\gamma_{\text{out}}$, i.e., $P_{\text{out}}=P_r[\gamma_{D} \leq \gamma_{\text{out}}]$, where $P_r[.]$ denotes the probability operation and $\gamma_{\text{out}}=(2^{2R}-1)$ with $R$ being a fixed spectral efficiency. The outage probability of the considered  proposed OVLC system is derived by substituting $\gamma_{{{\it \text{out}}}}$ in (\ref{CDF}) as follows
\begin{equation}
P_{\text{out}} = F_{\gamma_D}(\gamma_{\text{out}})
\end{equation}

%
%

\subsection{Ergodic Capacity}
The ergodic capacity can be expressed in terms of the PDF of $\gamma_i$ as \cite{al2017precise}
\begin{align} \label{erg_cap}
{C}=\frac{1}{\mathrm{ln}(2)}\int _{0}^{\infty }\mathrm{ln}(1+\gamma)f_{\gamma _{\sf D}}(\gamma) d\gamma. \end{align}

By substituting (\ref{PDF}) in (\ref{erg_cap}), we obtain a closed form expression for the ergodic capacity as
\begin{equation}
\begin{aligned}
C =&  \frac{\pi}{\ln (2)} \left( -R{{\sl K}_{\alpha-\beta}\left(\sqrt [4]{\it Ax}\right)}
\left( {{\sl K}_{\alpha-\beta}\left(\sqrt [4]{{\it Bx}}\right)} \right) ^{2} 
\csc \left( \frac{3(\alpha + \beta)\pi}{4}  \right)\right) \\
&  + \frac{\pi}{\ln (2)} \left( Q \left( {{\sl K}_{\alpha-\beta}\left(\sqrt [4]{{\it Bx}}\right)} \right) ^{2}\csc \left( \frac{(\alpha +  \beta)\pi}{2}  \right)\right) \\
& +  P{{\sl K}_{\alpha-\beta}\left(\sqrt [4]{{\it Ax}}\right)}
\csc \left(\frac{(\alpha + \beta)\pi}{4} \right),
\end{aligned} 
\end{equation}
\noindent
where
$
R = 64\,{\frac {\left( {\frac {\alpha \beta }{
				\sqrt {\bar{\gamma}_{{{\sf sr}}}}}} \right) ^{(\alpha+\beta)/2}\left( {\frac {\alpha \beta } {\sqrt {\bar{\gamma}_{{{\sf rd}}}}}} \right) ^{(\alpha+\beta)} }{ \left( 
		\alpha +\beta \right) ^{3} \left( \Gamma \left( \alpha \right)  \right) ^{3} \left(\Gamma \left( \beta \right)  \right) ^{3}}} , \ A = \left(2\,\sqrt {{\frac {\sqrt {\alpha \beta}}{\sqrt {\bar{\gamma}_{{{\sf sr}}}}}}}\right),
$
$
Q = 16\,{\frac {\left( {\frac {\alpha \beta } {\sqrt {\bar{\gamma}_{{{\sf rd}}}}}} \right) ^{(\alpha+\beta)} }{ \left( 
		\alpha +\beta \right) ^{3} \left( \Gamma \left( \alpha \right)  \right) ^{2} \left(\Gamma \left( \beta \right)  \right) ^{2}}} , \ B = \left(2\,\sqrt {{\frac {\sqrt {\alpha \beta}}{\sqrt {\bar{\gamma}_{{{\sf rd}}}}}}}\right),
$
$
P = 4\,{\frac {\left( {\frac {\alpha \beta }{
				\sqrt {\bar{\gamma}_{{{\sf sr}}}}}} \right) ^{(\alpha+\beta)/2} }{ \left( 
		\alpha +\beta \right) \left( \Gamma \left( \alpha \right)  \right) \left(\Gamma \left( \beta \right)  \right)}}.
$

\section{Simulations and Numerical Results}
\begin{figure}[ht] 
	\centering    
	\includegraphics[width=0.350\textwidth]{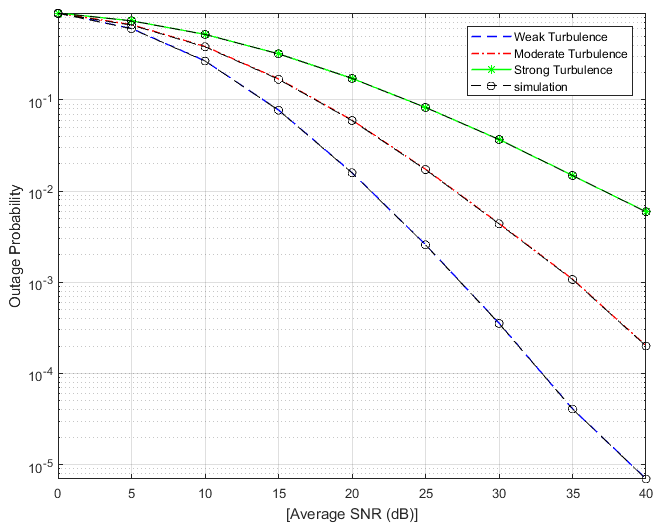}
	\caption[Outage Probability versus Average SNR for an Outdoor VLC Network]{Outage Probability versus average SNR for weak turbulence ($\alpha = 8.1, \beta = 4$), moderate turbulence ($\alpha = 4.2, \beta = 3$), and strong turbulence ($\alpha = 2.2, \beta = 2$)}
	\label{fig:outage}
\end{figure}

\begin{figure}[ht] 
	\centering    
	\includegraphics[width=0.350\textwidth]{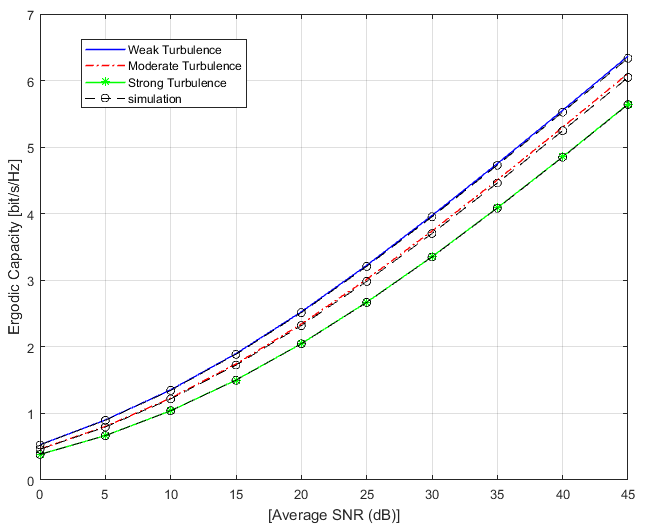}
	\caption[Ergodic Capacity versus Average SNR for an Outdoor VLC Network]{Ergodic Capacity versus average SNR for weak turbulence ($\alpha = 8.1, \beta = 4$), moderate turbulence ($\alpha = 4.2, \beta = 3$), and strong turbulence ($\alpha = 2.2, \beta = 2$)}
	\label{fig:erg_cap}
\end{figure}

In this section, we present a numerical evaluation of the outage probability and ergodic capacity to ascertain the achieved mathematical expressions obtained in the previous section. The numerical results are obtained using monte-carlo simulation. In each simulation, we consider different values of $\alpha$ and $\beta$, and different turbulence scenarios as used in \cite{al2017precise}. The outage probability and ergodic capacity performances for the OVLC system are given by Figures \ref{fig:outage} and \ref{fig:erg_cap}. Each figure also compares simulation results with numerical results from the derived relations for the outage probability and ergodic capacity. It observed that the performance of the weak turbulent channel is better than that of the moderate turbulence which also outperforms the strong turbulence scenario for both the ergodic capacity and the outage probability performance measures. The results indicate that the derived expressions can be used to evaluate the system performance without relying on time consuming simulations. 

\begin{figure}[ht] 
	\centering    
	\includegraphics[width=0.350\textwidth]{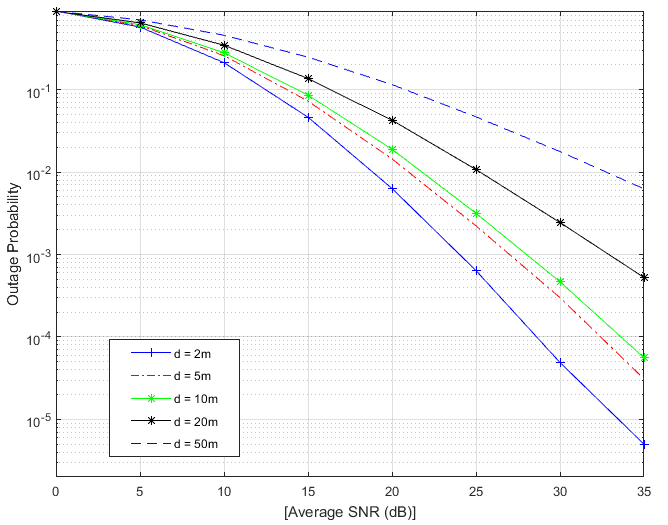}
	\caption[Ergodic Capacity versus Average SNR for an Outdoor VLC Network]{Outage probability versus average SNR for different Transmission distances between R and D for weak turbulence ($\alpha = 8.1, \beta = 4$)}
	\label{fig:outage_dis}
\end{figure}

\begin{figure}[ht] 
	\centering    
	\includegraphics[width=0.350\textwidth]{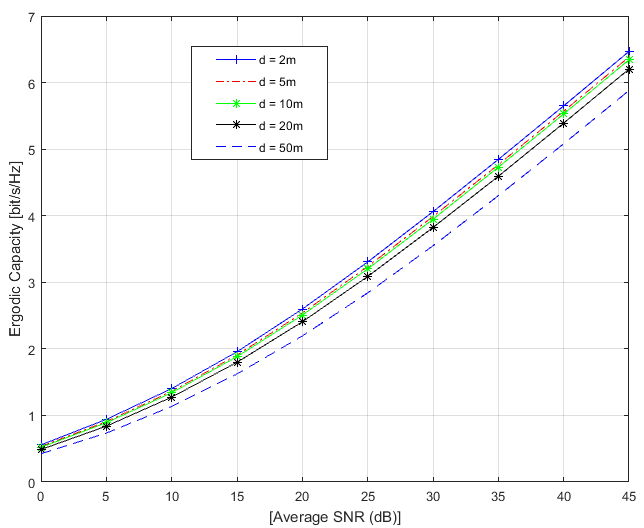}
	\caption[Ergodic capacity versus average SNR for an Outdoor VLC Network]{Ergodic capacity versus average SNR for different transmission distances between R and D for weak turbulence ($\alpha = 8.1, \beta = 4$)}
	\label{fig:cap_dis}
\end{figure}

To further examine the system performance, we fixed the values of $\alpha$ and $\beta$ at a weak turbulence condition and vary the distance between the two vehicles in Figures \ref{fig:outage_dis} and \ref{fig:cap_dis}. The results show that, the performance of the system degrades with increasing the distance,$d$ between the two vehicles in terms of both ergodic capacity and outage probability as shown in Figures \ref{fig:cap_dis} and \ref{fig:outage_dis} respectively. This is due to the fact that, as the distance between the transmitter (Vehicle 1) and receiver (Vehicle 2) increases, the average SNR reduces, and hence, the system performance reduces. Also, it is well-known that, path loss is a function of distance and hence, the larger the distance the worse the system performance.

\section{Conclusion}
This paper presented performance analyses in terms of outage probability and ergodic capacity for a  dual-hop relay outdoor visible light communication  (OVLC) system for IoV applications. Mathematical analysis was carried to define the link parameters of the link in the model. The statistics of the effective SNR at the destination node were  derived by obtaining novel expressions of its PDF and CDF. This was followed by closed-form derivations of the outage probability and ergodic capacity of the system. The performance of the system was then evaluated using Monte-Carlo simulations and compared to the theoretical numeric results. The results indicated that the derived expressions can be used to evaluate the system performance without relying on time consuming simulations.

\ifCLASSOPTIONcaptionsoff
  \newpage
\fi

\bibliographystyle{ieeetr}
\bibliography{ref}

\begin{thebibliography}{10}

\bibitem{cheng2015routing}
J.~Cheng, J.~Cheng, M.~Zhou, F.~Liu, S.~Gao, and C.~Liu, ``Routing in internet
  of vehicles: A review,'' {\em IEEE Transactions on Intelligent Transportation
  Systems}, vol.~16, no.~5, pp.~2339--2352, 2015.

\bibitem{hamid2019internet}
U.~Z.~A. Hamid, H.~Zamzuri, and D.~K. Limbu, ``Internet of vehicle (iov)
  applications in expediting the implementation of smart highway of autonomous
  vehicle: A survey,'' in {\em Performability in Internet of Things},
  pp.~137--157, Springer, 2019.

\bibitem{ndjiongue2018overview}
A.~Ndjiongue and H.~Ferreira, ``An overview of outdoor visible light
  communications,'' {\em Transactions on Emerging Telecommunications
  Technologies}, vol.~29, no.~7, p.~e3448, 2018.

\bibitem{lourencco2012visible}
N.~Louren{\c{c}}o, D.~Terra, N.~Kumar, L.~N. Alves, and R.~L. Aguiar, ``Visible
  light communication system for outdoor applications,'' in {\em Communication
  Systems, Networks \& Digital Signal Processing (CSNDSP), 2012 8th
  International Symposium on}, pp.~1--6, IEEE, 2012.

\bibitem{sun2017superimposed}
Z.-G. Sun, H.~Yu, and Y.-J. Zhu, ``A superimposed relaying strategy and power
  allocation for outdoor visible light communications,'' {\em IEEE Access},
  vol.~5, pp.~9555--9561, 2017.

\bibitem{lee2009performance}
I.~E. Lee, M.~L. Sim, and F.~W.-L. Kung, ``Performance enhancement of outdoor
  visible-light communication system using selective combining receiver,'' {\em
  IET optoelectronics}, vol.~3, no.~1, pp.~30--39, 2009.

\bibitem{kim2012outdoor}
D.-R. Kim, S.-H. Yang, H.-S. Kim, Y.-H. Son, and S.-K. Han, ``Outdoor visible
  light communication for inter-vehicle communication using controller area
  network,'' in {\em Communications and Electronics (ICCE), 2012 Fourth
  International Conference on}, pp.~31--34, IEEE, 2012.

\bibitem{luo2015performance}
P.~Luo, Z.~Ghassemlooy, H.~Le~Minh, E.~Bentley, A.~Burton, and X.~Tang,
  ``Performance analysis of a car-to-car visible light communication system,''
  {\em Applied Optics}, vol.~54, no.~7, pp.~1696--1706, 2015.

\bibitem{lin2017outage}
S.-H. Lin, J.-Y. Wang, X.~Bao, and Y.~Li, ``Outage performance analysis for
  outdoor vehicular visible light communications,'' in {\em Wireless
  Communications and Signal Processing (WCSP), 2017 9th International
  Conference on}, pp.~1--5, IEEE, 2017.

\bibitem{salhab2016power}
A.~M. Salhab, F.~S. Al-Qahtani, R.~M. Radaydeh, S.~A. Zummo, and H.~Alnuweiri,
  ``Power allocation and performance of multiuser mixed rf/fso relay networks
  with opportunistic scheduling and outdated channel information,'' {\em
  Journal of Lightwave Technology}, vol.~34, no.~13, pp.~3259--3272, 2016.

\bibitem{korenev2003bessel}
B.~G. Korenev, {\em Bessel functions and their applications}.
\newblock CRC Press, 2003.

\bibitem{al2017precise}
O.~M.~S. Al-Ebraheemy, A.~M. Salhab, A.~Chaaban, S.~A. Zummo, and M.-S.
  Alouini, ``Precise outage analysis of mixed rf/unified-fso df relaying with
  hd and 2 im-dd channel models,'' in {\em Wireless Communications and Mobile
  Computing Conference (IWCMC), 2017 13th International}, pp.~1184--1189, IEEE,
  2017.

\end{thebibliography}


\end{document}